\documentclass[conference]{IEEEtran}
\usepackage{cite}
\usepackage{amsmath,amssymb,amsfonts}
\usepackage{graphicx}
\usepackage{booktabs}
\usepackage{array}
\usepackage{url}
\usepackage{xcolor}
\usepackage{tikz}
\usetikzlibrary{positioning,arrows.meta,shapes.geometric,calc}
\usepackage{balance}
\newcolumntype{L}[1]{>{\raggedright\arraybackslash}p{#1}}

\newcommand{\RA}{R_{\!A}}
\newcommand{\Dphi}{\Delta\phi}
\newcommand{\Gstar}{G^{*}}

\newcommand{\exppanel}[3]{\fcolorbox{black!45}{black!3}{\begin{minipage}[c][#1][c]{0.9\columnwidth}\centering
\textbf{#2}\\[3pt]\footnotesize #3\end{minipage}}}
 
\begin{document}
\title{Contact-Free Nuclear and Cytoplasmic Mechanophenotyping by Amplitude-Modulated Ultrasound and Phase-Locked Optical Diffraction Tomography: Computational Feasibility and Initial Experimental Readouts}
\author{\IEEEauthorblockN{1\textsuperscript{st} Yagmur ISIK\textsuperscript{1},
2\textsuperscript{nd} Merve SEVGI\textsuperscript{2},
3\textsuperscript{rd} Esmahan CAGLAR\textsuperscript{2},
4\textsuperscript{th} Aysenur AKGUN\textsuperscript{3},\\
5\textsuperscript{th} Hasan Berkay ABDIOGLU\textsuperscript{1},
6\textsuperscript{th} Gokhan Bora ESMER\textsuperscript{4},
7\textsuperscript{th} Huseyin UVET\textsuperscript{1}\vspace{0.5em}}
\IEEEauthorblockA{\itshape
\textsuperscript{1}Department of Mechatronics Engineering, Yildiz Technical University, Istanbul, Turkey\\
\textsuperscript{2}Department of Bioengineering, Yildiz Technical University, Istanbul, Turkey\\
\textsuperscript{3}Department of Molecular Biology and Genetics, Yildiz Technical University, Istanbul, Turkey\\
\textsuperscript{4}Department of Control and Automation Engineering, Yildiz Technical University, Istanbul, Turkey}}
\maketitle

\begin{abstract}
We investigate whether acoustic stimulation combined with phase-locked optical diffraction tomography (ODT) could measure the mechanical responses of the nucleus and cytoplasm separately in adherent cells. A megahertz ultrasound carrier is amplitude-modulated at low frequencies, while angle-scanned holography provides a route to reconstruct the first-harmonic refractive-index response in three dimensions. We built a computational model connecting viscoelastic finite-element simulations of MDA-MB-231 cells with a refractive-index phantom, hologram formation, camera noise and ODT reconstruction. Under an assumed 0.1 Pa harmonic shear traction at 10 Hz, the model predicts 3.6 to 9.2 nm of nuclear shear and 6.6 to 15.1 nm of apical motion. The nuclear-to-cytoplasmic amplitude ratio $\RA$ is primarily sensitive to stiffness contrast in the tested scenarios, while the relative phase $\Dphi$ responds to differences in dissipation. A simulated protocol using 100 illumination angles and 16 phase visits per angle recovered an imposed stiffening and recovery trajectory after averaging four acquisitions. We also obtained initial 3-D tomograms of adherent MDA-MB-231 cells and report preliminary regional measurements from five cells. Their mean amplitude ratio was $0.616\pm0.098$, and their mean relative phase was $-1.91^\circ\pm0.58^\circ$ (mean and between-cell sample standard deviation). All five amplitude ratios were below unity, and all phase differences were negative, consistent with the qualitative directions of selected model scenarios. The small cohort and the absence of independently quantified experimental phase noise prevent a claim of quantitative mechanical validation. Together, these results support further evaluation of phase-locked acoustic ODT and identify experimental requirements for a controlled validation study.
\end{abstract}

\section{Introduction}
The nucleus and cytoplasm are mechanically distinct, but they respond together when a cell is deformed. Their stiffness and dissipation influence cell migration, force transmission and adaptation to mechanical loading. Measuring how each compartment responds in an intact adherent cell could help distinguish these effects. Repeated measurements on the same cell might also reveal how its mechanical state changes after a challenge. We refer to this proposed longitudinal response measure as mechanical adaptive reserve (MAR).

Established techniques capture parts of this picture. Atomic force microscopy and intracellular microrheology report apparent stiffness and viscosity of breast-cancer cells, including a comparatively compliant, viscoelastic cytoplasm and a stiffer nuclear region \cite{Fischer2020,Dessard2024,Rother2014,Zbiral2023}, but values for the same cell line differ widely between methods \cite{Wu2018}, and contact-based probes interrogate one location at a time. Optical approaches avoid contact: optical diffraction tomography (ODT) reconstructs the three-dimensional (3-D) refractive-index (RI) distribution of living cells label-free \cite{Wolf1969,Sung2009,Park2018}, and combined Brillouin-ODT microscopy has resolved compartment-specific longitudinal moduli at gigahertz frequencies \cite{Schlussler2022}. What is missing is a way to drive a cell at physiologically relevant, low frequencies without contact and to read out the resulting nuclear and cytoplasmic deformation in 3-D.

Acoustic stimulation combined with holographic imaging is already established for contact-free cellular stiffness mapping. Varol and colleagues reconstructed surface stiffness maps from acoustically driven deformations measured by digital holographic microscopy, with reference-particle and atomic force microscopy comparisons \cite{Varol2022}. The present study addresses a different measurement target: the amplitude and phase of nuclear and cytoplasmic responses within a three-dimensional RI volume, rather than a stiffness map inferred from surface motion. We consider an 8 MHz ultrasound carrier with a 10 Hz amplitude envelope, for which acoustic radiation forces and streaming may provide a time-varying load \cite{Gorkov1962,Bruus2012,Imashiro2022}. Phase-referenced angular holography is then used to investigate the corresponding harmonic optical signal. A reliable conversion of this signal into mechanical strain remains a central part of the proposed inverse problem.

We test the idea with a viscoelastic model of adherent MDA-MB-231 cells, a simulated ODT acquisition and camera-level holographic reconstruction. The analysis considers how relative nuclear and cytoplasmic amplitudes and phases depend on material properties, how missing-angle information and noise affect recovery, and whether a prescribed stiffening and recovery trajectory can be tracked. We then describe the experimental platform, initial cellular tomograms and preliminary measurements from five cells. The experimental observations are reported independently of the illustrative simulation cases (Section~\ref{sec:exp}).

\section{Framework and Methods}
\subsection{Acoustic-ODT measurement concept}
\begin{figure}[t]
\centering
\IfFileExists{figures/F0_cell_concept.jpg}{%
  \includegraphics[width=\columnwidth]{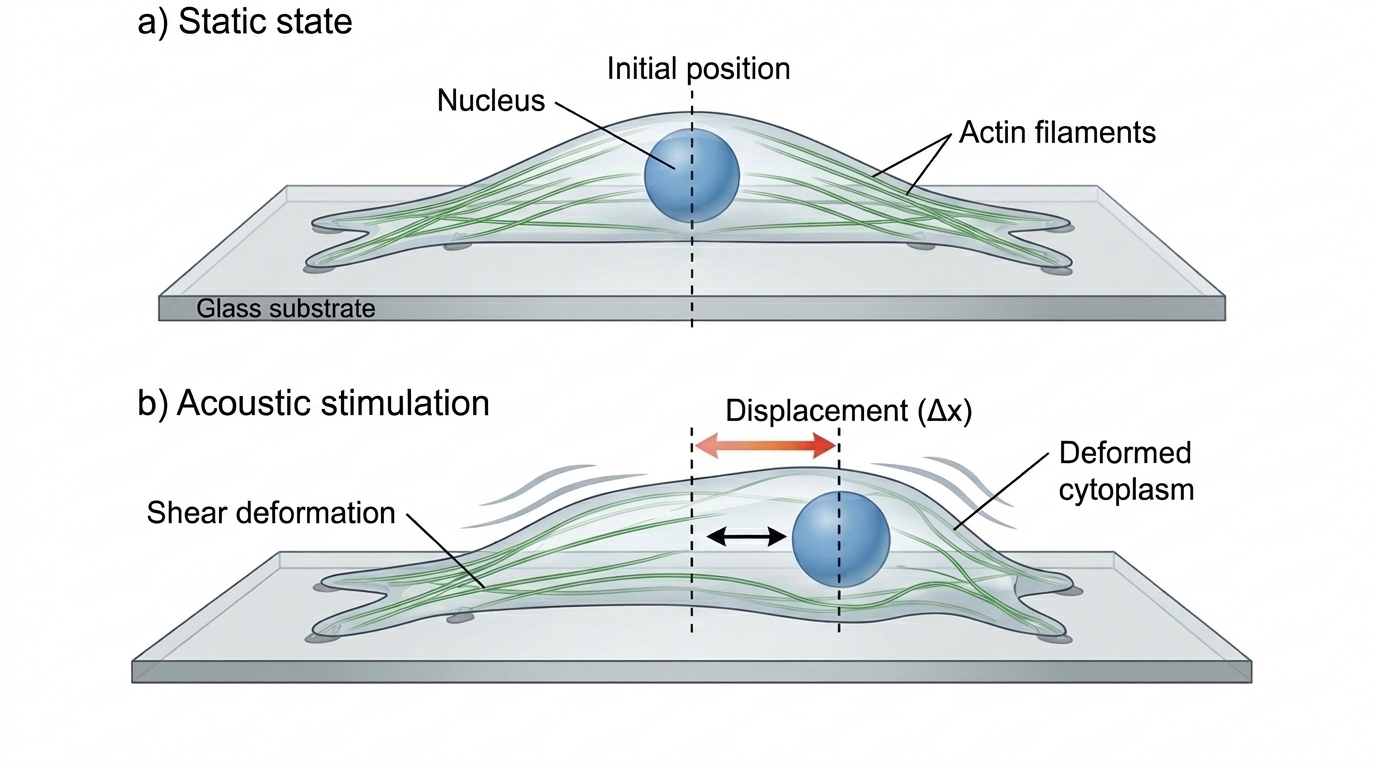}}{%
  \exppanel{2.3cm}{Figure: Measurement principle}{Adherent cell at rest (a) and under\\amplitude-modulated acoustic stimulation (b).}}
\caption{Measurement principle. (a) An adherent cell at rest on the glass substrate, with the nucleus embedded in the actin-reinforced cytoplasm. (b) Under amplitude-modulated acoustic stimulation, near-surface streaming exerts a harmonic lateral traction that shears the cytoplasm and displaces the nucleus. Because the nucleus differs from the cytoplasm in stiffness and dissipation, the two compartments respond with different amplitudes and phases, quantified by $\RA$ and $\Dphi$. Deformations are exaggerated for clarity; predicted amplitudes are a few nanometers.}
\label{fig:concept}
\end{figure}

\begin{figure*}[t]
\centering
\IfFileExists{figures/F0_phase_resolved_workflow.png}{%
  \includegraphics[width=\textwidth]{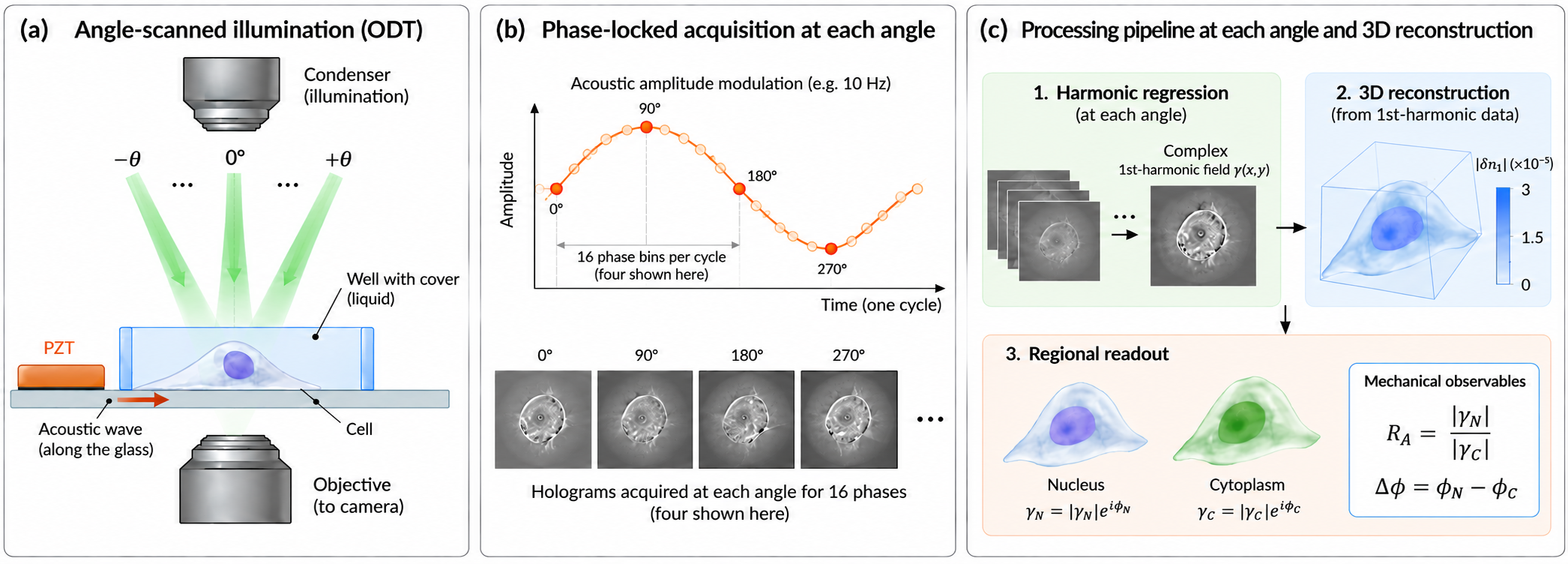}}{%
  \fbox{\parbox[c][3cm][c]{0.9\textwidth}{\centering Figure: phase-resolved acousto-optic measurement}}}
\caption{Phase-resolved acousto-optic measurement. (a) The acoustic wave launched by the amplitude-modulated PZT travels along the coverslip into the well, while the DMD steps the illumination angle through the condenser; the objective below the coverslip collects the transmitted field. (b) At each angle, holograms are recorded at 16 phases of the 10 Hz modulation (four shown). (c) Harmonic regression of the phase series at each angle yields the complex first-harmonic field; a single reconstruction gives the first-harmonic RI map, from which the regional amplitude ratio $\RA$ and phase difference $\Dphi$ are obtained.}
\label{fig:workflow}
\end{figure*}

Figure~\ref{fig:concept} illustrates the mechanical principle and Fig.~\ref{fig:workflow} the measurement workflow. An RF carrier near 8~MHz is amplitude-modulated (reference modulation 10 Hz; sweep 5 to 200 Hz). The first harmonic of the resulting load drives a complex displacement field in the cell, which phase-locked, angle-scanned holography converts into a first-harmonic RI map. Regional fitting of this map yields the nuclear and cytoplasmic mechanical responses.

\subsection{Viscoelastic cell model}
A dome-shaped cell adheres to a glass plane (base radius 16 $\mu$m, apical height 6.5 $\mu$m, meshed volume $\approx$2790 $\mu$m$^3$, consistent with the 17.3 $\mu$m suspended diameter of MDA-MB-231 cells \cite{Yang2016}). An oblate nucleus (13 $\times$ 13 $\times$ 4.5 $\mu$m, $\approx$425 $\mu$m$^3$ in the mesh) sits 3.25 $\mu$m above the substrate; an optional superficial layer represents an effective cortex. All compartments are linear viscoelastic with frequency-dependent complex shear moduli $\Gstar(\Omega)=G'(\Omega)+iG''(\Omega)$ and $\nu=0.49$, reflecting undrained behavior at 10 Hz; harmonic quantities follow $x(t)=\mathrm{Re}\{\hat{x}e^{i\Omega t}\}$. Table~\ref{tab:params} lists the inputs and their basis.

Small-strain harmonic equilibrium is solved for each load,
\begin{equation}
\nabla\!\cdot\hat{\boldsymbol\sigma}+\hat{\mathbf f}=0,\quad
\hat{\boldsymbol\sigma}=2G^*\hat{\boldsymbol\varepsilon}
+\lambda^*\operatorname{tr}(\hat{\boldsymbol\varepsilon})\mathbf I,
\label{eq:mechanics}
\end{equation}
with $\hat{\boldsymbol\varepsilon}=\tfrac12(\nabla\hat{\mathbf u}+\nabla\hat{\mathbf u}^{\mathsf T})$, using quadratic hexahedral elements in scikit-fem. A quarter-symmetric domain with load-specific symmetry conditions reproduces the full-domain solution to numerical precision and reduces the system to $\approx$11{,}300 degrees of freedom; a refined mesh ($\approx$22{,}500) changes the reported observables by less than 7\%.

\begin{table}[t]
\caption{Model inputs at 10 Hz and their basis}
\label{tab:params}
\centering
\footnotesize
\setlength{\tabcolsep}{3pt}
\begin{tabular}{L{0.30\columnwidth}L{0.25\columnwidth}L{0.35\columnwidth}}
\toprule
Quantity & Value & Basis \\
\midrule
Cytoplasm $G_C^*$, reference & $38.5+2.2i$ Pa & Maxwell, $G=38.6$ Pa, $\eta=10.7$ Pa\,s \cite{Dessard2024} \\
Cytoplasm $G_C^*$, alternative & $88.3+2.6i$ Pa & generalized Maxwell \cite{Zbiral2023} \\
Stiffness ratio $r_G=|G_N^*|/|G_C^*|$ & 1.5 (scan 1 to 10) & apparent nuclear vs.\ cytoskeletal moduli \cite{Fischer2020} \\
Nuclear loss tangent & 0.2 (scan 0.05 to 0.5) & sensitivity parameter \\
Cortex (optional) & $\approx1.6+1.1i$ kPa, $\sim$1 $\mu$m & whole-cell dynamic AFM \cite{Rother2014} \\
Acoustic contrast $\Phi$ & 0.03 & compressibility \cite{Yang2016} \\
Loads & 100 J/m$^3$; 0.1 Pa & scenarios (linear scaling) \\
\bottomrule
\end{tabular}
\end{table}

\subsection{Acoustic loading}
The transducer drive is
\begin{equation}
p(t)=p_0\sqrt{\frac{1+\cos(\Omega t)}{2}}\cos(\omega_ct),
\label{eq:drive}
\end{equation}
with $\omega_c/2\pi=8$ MHz and $\Omega/2\pi=10$ Hz. The square-root envelope makes the time-averaged acoustic energy density, and hence the force, contain only a static term and a pure first harmonic at $\Omega$.

Two loading mechanisms are evaluated. For a standing field near the glass, the radiation-force density on the cell is
\begin{equation}
\hat{f}_z(z)=\frac{3}{2}\Phi k E_{\mathrm{ac}}\sin(2kz),
\label{eq:radiation}
\end{equation}
where $k$ is the acoustic wavenumber and $E_{\mathrm{ac}}=100$ J/m$^3$ the peak energy density; the prefactor is half that of the steady force density because the first harmonic of the envelope carries half of its peak. The resulting first-harmonic force on the whole cell is 59 pN at 8 MHz. Acoustic streaming is represented by a uniform lateral harmonic traction $\hat t_x=0.1$ Pa on the apical surface. Both magnitudes are scenarios that scale linearly; their actual values in the device are to be calibrated experimentally (Section~\ref{sec:exp}).

\subsection{Regional observables and mechanical adaptive reserve}
From the complex engineering shear strain $\hat\gamma_{xz}=\partial\hat u_x/\partial z+\partial\hat u_z/\partial x$, averaged over the nucleus ($N$) and the surrounding cytoplasm ($C$),
\begin{align}
\RA &= \frac{|\overline{\hat\gamma}_{N}|}{|\overline{\hat\gamma}_{C}|},\qquad
\Dphi = \arg(\overline{\hat\gamma}_{N})-\arg(\overline{\hat\gamma}_{C}).
\label{eq:metrics}
\end{align}
A negative $\Dphi$ indicates that the nucleus lags the cytoplasm. Being ratios and differences, both observables are insensitive to the absolute load. To quantify adaptation after a mechanical challenge, we use
\begin{align}
A &= \RA(t_{\mathrm{pre}})-\RA(t_{\mathrm{post}}),\qquad \mathrm{MAR} = A\rho,\\
\rho &= \frac{\RA(t_{\mathrm{last}})-\RA(t_{\mathrm{post}})}{A},
\label{eq:mar}
\end{align}
where $A$ describes the initial change in $\RA$ and $\rho$ the recovered fraction, defined for nonzero $A$. By construction, $\mathrm{MAR}=\RA(t_{\mathrm{last}})-\RA(t_{\mathrm{post}})$. MAR is therefore a descriptive recovery difference, not an independent mechanical parameter. The imposed trajectory is used to test numerical tracking, not to establish biological adaptation.

\subsection{Synthetic optical acquisition and reconstruction}
The static RI phantom ($n_m=1.337$, $n_C=1.366$, $n_N=1.362$) carries spatially correlated, organelle-like texture on a 0.1 $\times$ 0.1 $\times$ 0.2 $\mu$m grid. To first order, motion advects RI contrast,
\begin{equation}
\delta\hat n(\mathbf r)\simeq-\hat{\mathbf u}(\mathbf r)\cdot\nabla n_0(\mathbf r).
\label{eq:ri}
\end{equation}
Two levels of optical realism are used. The image-space model filters $\delta\hat n$ with the angle-scanned ODT transfer function, including the missing cone \cite{Lim2015}, and adds band-limited noise (100 independent realizations per condition). The camera-level model generates Rytov fields at 100 illumination angles (532 nm, assumed illumination NA 1.15 and detection NA 1.2) using Fourier-diffraction mapping \cite{Wolf1969}, forms off-axis holograms, and adds shot noise, 3 e$^-$ read noise, 8-bit quantization, random frame-wise phase offsets and 20 $\mu$s trigger jitter for an IMX421-class sensor (24{,}600 e$^-$ full well and 60\% fill). These are simulation parameters, not measured performance limits of the assembled instrument.

The phase-resolved acquisition is interleaved rather than completed angle by angle. Each DMD sweep visits all 100 angles once, and 16 sweeps provide 16 frames per angle. The simulated camera operates at approximately 301.9 frames/s, so the revisit interval for a fixed angle is $100/301.9\approx0.331$ s. At 10 Hz this corresponds to three full periods plus approximately $5/16$ of a period. The relative phase-bin sequence therefore advances by five positions modulo 16 and covers all bins before repeating. The 1600-frame sequence lasts approximately 5.3 s. The phase tags must be determined from the frame timestamps and the common modulation reference; neither sequential phase-bin order nor identical absolute phase offsets for all angles is assumed. This timing calculation applies to the 100-angle simulated acquisition. A 200-angle phase-resolved acquisition requires its own timing validation.

The simulated holograms are first demodulated spatially by Fourier-sideband filtering to retrieve a complex field at each frame. At each illumination angle, a reference-visit field ratio is formed and corrected using a cell-free background region to suppress global phase fluctuations. The complex logarithm of this corrected field ratio is then taken before fitting its time dependence. The regression includes a constant term, linear drift and sine/cosine coefficients at the first and second modulation harmonics. Its first-harmonic Rytov coefficients are mapped into 3-D Fourier space and reconstructed once. The order matters: taking the logarithm of a demodulated field harmonic is not equivalent to demodulating the logarithm. In the weak-signal limit only, the Rytov harmonic can also be approximated as $\chi_1\simeq E_1/E_0$.

In each compartment, the recovered harmonic RI signal is fitted with a complex affine displacement model. With $\hat{\mathbf u}(\mathbf r)=\hat{\mathbf A}(\mathbf r-\mathbf r_c)+\hat{\mathbf b}$ and the linearized advection relation in (\ref{eq:ri}), the sampled RI changes form a linear inverse problem. The implemented fit includes a separate affine nuisance baseline and estimates shear from $\hat A_{xz}+\hat A_{zx}$. Multiple voxels constrain the unknown coefficients, but do not guarantee a unique or stable solution. Low RI gradients, similar gradient directions, segmentation errors and the missing cone can degrade the rank and conditioning of the design matrix. The reported simulations test recovery under selected phantoms and noise levels, rather than establishing general identifiability across arbitrary cell structures. Without camera noise, the camera-level chain reproduces the image-space result with an agreement coefficient of 0.993.

\subsection{Experimental platform}
\label{sec:platform}
Figure~\ref{fig:setup} shows the acousto-optic chip. 

A thickness-mode PZT plate (CTS 3195HD, 38.5 $\times$ 7.5 $\times$ 0.25 mm, nominal resonance 8.44 MHz) is bonded to the upper surface of a 24 $\times$ 60 mm coverslip. A protective cover maintains an air gap above the PZT and the acoustic propagation path to minimize mechanical loading from the housing. The amplitude-modulated drive signal is generated by an arbitrary waveform generator (Siglent SDG6032X) and amplified by a high-voltage amplifier (Falco Systems WMA-300). The acoustic wave propagates along the glass substrate toward a $10\times10$ mm well containing cells immersed in buffer. The well is sealed with a glass lid to stabilize the optical path. Polyacrylamide (PAA) microbeads are intended for flow tracking and mechanical reference measurements.

The chip is mounted on an ODT microscope equipped with DMD-based angular illumination scanning and off-axis holographic detection (Fig.~\ref{fig:odt}). Two 100$\times$ oil-immersion objectives serve as the condenser and imaging objective, respectively. In the aqueous sample, the maximum propagating numerical aperture is bounded by the refractive index of the medium ($\approx$1.33). Camera exposures, DMD pattern updates, and the ultrasound drive are synchronized using a common timing reference. Holograms are recorded using a global-shutter camera (Basler boA1936-400cm).

\subsection{Cell Culture and Sample Preparation}

MDA-MB-231 cells were obtained from the American Type Culture Collection (ATCC) and maintained in DMEM/F12 medium supplemented with 10\% fetal bovine serum (FBS) at $37\,^{\circ}\mathrm{C}$ in a humidified atmosphere containing 5\% CO$_2$.

Approximately 1,000 cells were seeded onto glass coverslips and cultured for 24 h before optical diffraction tomography (ODT) imaging. During ODT acquisition, cells were maintained in buffer at room temperature.

\subsection{Preliminary regional measurements}
Five adherent MDA-MB-231 cells were examined using an 8 MHz ultrasound carrier amplitude-modulated at 10 Hz. The standard DMD acquisition used 100 illumination angles, matching the angular sampling adopted in the computational study. Some recordings were extended to 200 angles to obtain denser angular sampling. Hardware triggers coordinated camera exposures, DMD pattern updates and the ultrasound generator, so that the recorded frames could be referenced to the 10 Hz modulation cycle. The phase-resolved protocol was designed around 16 phase visits at each illumination angle.

The proposed signal-processing sequence starts with Fourier-sideband recovery of the complex optical field from each hologram. Phase-tagged acquisitions allow temporal harmonic regression, with constant, drift, first-harmonic and second-harmonic terms. As demonstrated in the computational workflow, this is applied to the complex logarithm of a background-corrected field ratio before a single first-harmonic ODT reconstruction. Nuclear and cytoplasmic responses are then estimated by separate regional affine fits, yielding $\RA$ and $\Dphi$ according to (\ref{eq:metrics}). These are the model's processing steps; the five-cell experimental results remain preliminary until their exact frame-level timing, fit residuals and instrumental uncertainty are documented.

The five experimental measurements are reported individually and as a mean and between-cell sample standard deviation. They are not paired with specific simulated cells, so we compare only qualitative patterns. Independent-acquisition repeatability, ultrasound-off controls and a measured phase-noise floor are not yet available. Although trigger synchronization was used, a shared hardware trigger alone does not quantify the residual phase error. The small measured phase differences must therefore be regarded as exploratory observations rather than validated measurements of nuclear dissipation.

\begin{figure}[t]
\centering
\IfFileExists{figures/F0_chip_top_side.jpeg}{%
  \includegraphics[width=0.78\columnwidth]{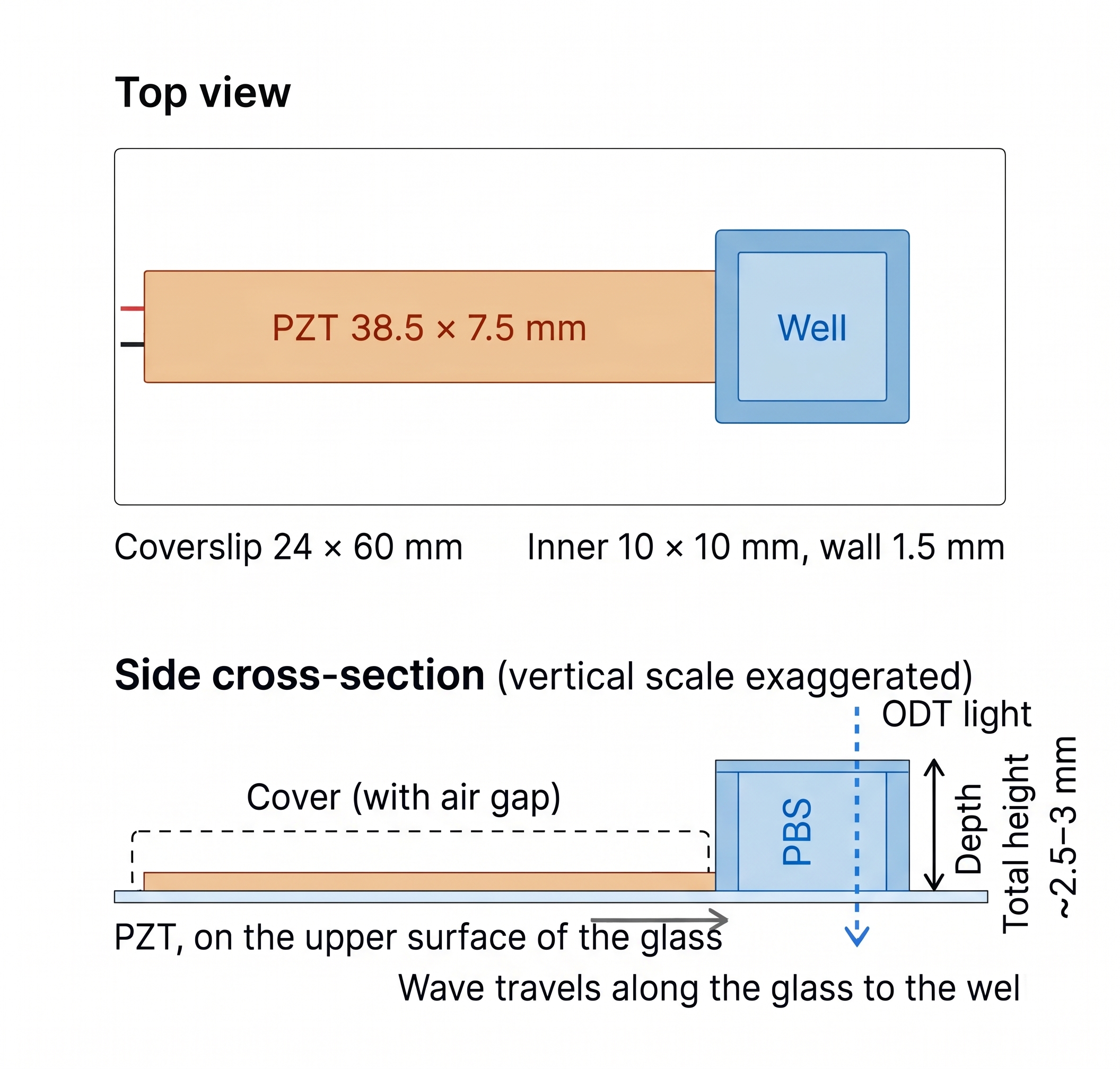}}{%
  \exppanel{2.6cm}{Figure: Acousto-optic chip}{Top view and side cross-section:\\PZT on coverslip, air-gap cover, lidded well.}}
\caption{Acousto-optic chip, top view and side cross-section (vertical scale exaggerated). A $38.5\times7.5$ mm thickness-mode PZT plate is bonded to the upper surface of a $24\times60$ mm coverslip and protected by a cover that leaves an air gap above the plate and the propagation path. The amplitude-modulated wave travels along the glass into a $10\times10$ mm well (wall thickness 1.5 mm) filled with PBS and closed by a glass lid, which stabilizes the optical path of the vertically transmitted ODT illumination.}
\label{fig:setup}
\end{figure}

\begin{figure}[t]
\centering
\IfFileExists{figures/F0_odt_optical_setup.jpg}{%
  \includegraphics[width=\columnwidth]{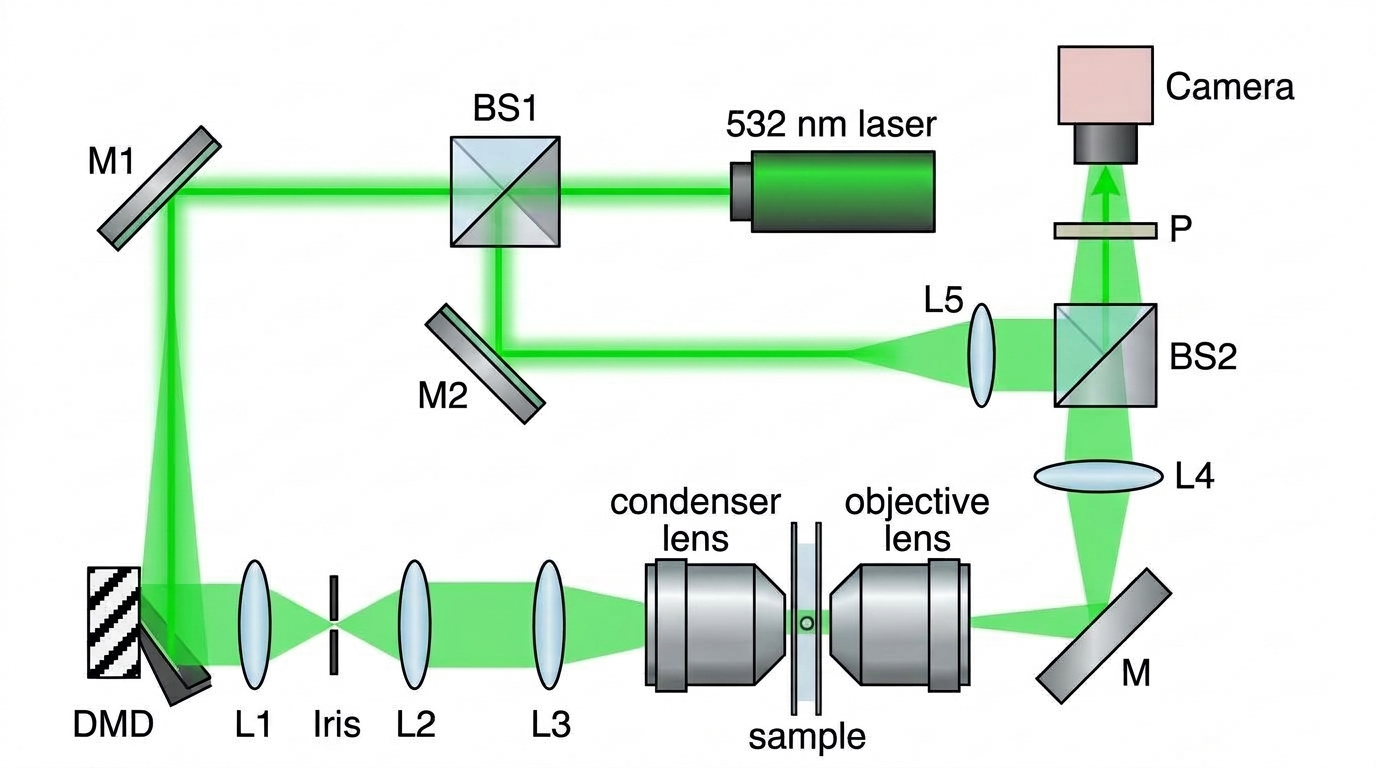}}{%
  \exppanel{2.3cm}{Figure: ODT optical setup}{DMD-based angle scanning,\\off-axis holographic detection.}}
\caption{Optical diffraction tomography setup. The 532 nm laser beam is divided by a beam splitter (BS1) into a sample and a reference arm. The illumination angle of the plane waves impinging on the sample is controlled by a digital micromirror device (DMD) and the iris. Condenser: infinity-corrected oil-immersion objective (100$\times$, NA 1.25); objective: oil-immersion objective (100$\times$, NA 1.25). L1-L5: lenses; M, M1, M2: mirrors; BS1, BS2: beam splitters.}
\label{fig:odt}
\end{figure}

\section{Results: Computational Feasibility}
\subsection{Nanometer-scale nuclear and cytoplasmic deformation}
Under a 0.1 Pa lateral traction at 10 Hz, the soft-cell scenarios produce 3.6 to 9.2 nm of nuclear shear and 6.6 to 15.1 nm of apical lateral motion (Table~\ref{tab:mechanics}). The radiation-force scenario at 8 MHz and 100 J/m$^3$ yields a 1.0 to 2.3 nm change in lateral nuclear dimension; shear loading is therefore the more efficient drive for flattened adherent cells, and it moves the cell in-plane, where ODT resolution is highest. Shear results do not depend on the carrier, whereas radiation results scale with $f_c^2$ (59 pN at 8 MHz versus 92 pN at 10 MHz). A cytoplasm as stiff as 1 kPa reduces all responses below 1 nm, which makes the cytoplasmic modulus the main determinant of signal strength.

\begin{table}[t]
\caption{Simulated deformation at 8 MHz carrier and 10 Hz modulation}
\label{tab:mechanics}
\centering
\footnotesize
\setlength{\tabcolsep}{3.0pt}
\begin{tabular}{lrrrr}
\toprule
Scenario & $u_{\rm apex}^{\rm shear}$ & $d_{N}^{\rm shear}$ & $d_N^{\rm rad}$ & $\RA$ \\
& (nm) & (nm) & (nm) & \\
\midrule
A: reference & 15.14 & 9.18 & 2.28 & 0.769 \\
B: A + cortex & 7.46 & 5.30 & 1.29 & 0.699 \\
C: 88 Pa cytoplasm & 6.60 & 4.01 & 1.00 & 0.769 \\
D: $r_G=5$ & 10.98 & 3.64 & 1.12 & 0.333 \\
\bottomrule
\end{tabular}
\vspace{2pt}
\parbox{0.96\columnwidth}{\scriptsize Shear columns: 0.1 Pa lateral traction; radiation column: 100 J/m$^3$ peak energy density. $d_N^{\rm shear}=|\overline{\hat\gamma}_N|\times4.5\,\mu$m; $d_N^{\rm rad}=|\overline{\hat\varepsilon}_{xx,N}|\times13\,\mu$m.}
\end{table}

\subsection{Amplitude ratio and phase separate stiffness from damping}
\begin{figure}[t]
\centering
\includegraphics[width=\columnwidth]{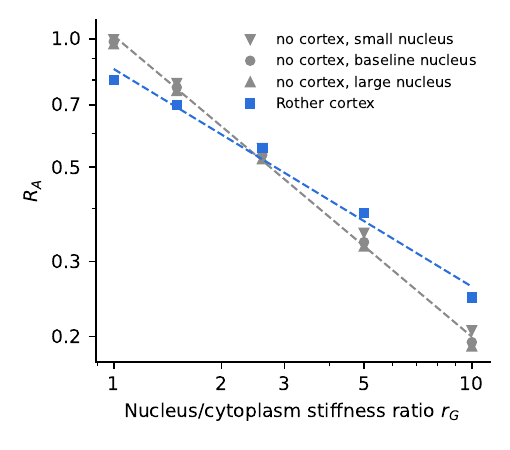}
\caption{Nuclear-to-cytoplasmic shear-amplitude ratio versus stiffness contrast $r_G$ for three nuclear sizes without cortex (grey) and with an effective cortical layer (blue). Dashed lines: power-law fits.}
\label{fig:calib}
\end{figure}

The amplitude ratio follows the stiffness contrast monotonically: $\RA\approx1.016\,r_G^{-0.707}$ without and $\RA\approx0.849\,r_G^{-0.511}$ with the cortical layer (Fig.~\ref{fig:calib}). Without cortex, a twofold change in nuclear volume (244 to 552 $\mu$m$^3$) alters the inferred $r_G$ by only $\pm$10\%, so $\RA$ is robust to cell-to-cell variation in nuclear size.

\begin{figure*}[t]
\centering
\includegraphics[width=0.86\textwidth]{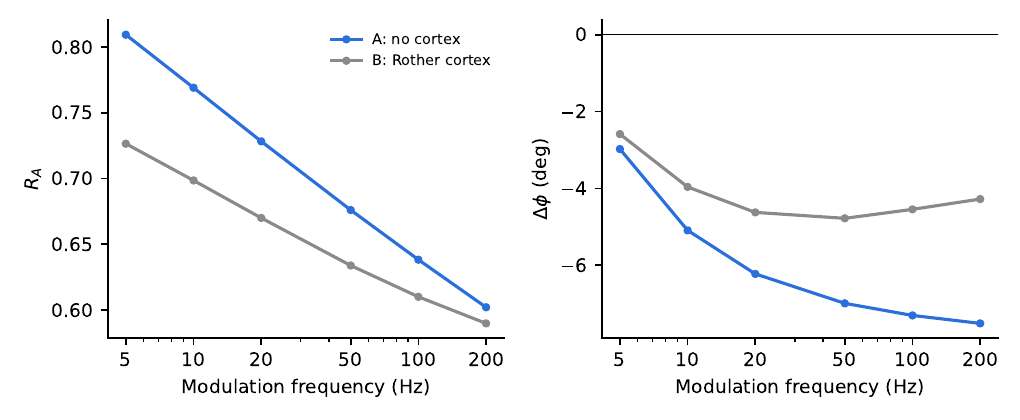}
\caption{Nuclear-to-cytoplasmic amplitude ratio (left) and phase difference (right) versus modulation frequency for the reference cell and with an effective cortex (8 MHz carrier, 0.1 Pa lateral traction).}
\label{fig:freq}
\end{figure*}

Varying stiffness and damping independently shows that the two observables are nearly orthogonal (Table~\ref{tab:separation}). A tenfold change in the nuclear loss tangent alters $\RA$ by less than 1\%, while $\Dphi$ is close to zero when both compartments dissipate equally and grows to $-15^\circ$ to $-18^\circ$ for a strongly dissipative nucleus. With the cortical layer the phase contrast retains 65 to 80\% of these values. Across modulation frequencies of 5 to 200 Hz, $\RA$ declines smoothly as the nucleus stiffens relative to the elastic plateau of the cytoplasm, and $\Dphi$ saturates near 50 Hz (Fig.~\ref{fig:freq}); the shape of these curves therefore also discriminates between rheological models.

\begin{table}[t]
\caption{Independent stiffness and damping contrast (no cortex)}
\label{tab:separation}
\centering
\footnotesize
\begin{tabular}{ccccc}
\toprule
$r_G$ & $\tan\delta_N$ & $\tan\delta_N/\tan\delta_C$ & $\RA$ & $\Dphi$ ($^\circ$) \\
\midrule
1.5 & 0.05 & 0.9 & 0.768 & $+0.3$ \\
1.5 & 0.2 & 3.5 & 0.769 & $-5.1$ \\
1.5 & 0.5 & 8.7 & 0.775 & $-14.8$ \\
5 & 0.05 & 0.9 & 0.333 & $+0.3$ \\
5 & 0.2 & 3.5 & 0.333 & $-6.0$ \\
5 & 0.5 & 8.7 & 0.335 & $-17.5$ \\
\bottomrule
\end{tabular}
\end{table}

\subsection{ODT recovers the mechanical contrast with a calibratable bias}
The image-space inverse test preserves the ordering of all scenarios. Without noise, it returns $\RA=0.86$ for a reference of 0.81 (S1, $r_G=1.5$) and 0.51 for 0.40 (S2, $r_G=5$), and $\Dphi=-8.2^\circ$ for a reference of $-12.7^\circ$ (S3, high nuclear loss). These deviations are systematic and consistent with the limited axial transfer of the missing cone, which attenuates the $\partial u_x/\partial z$ component of the shear strain. Their predictability in these examples suggests that a matched optical forward model may reduce the bias. A linear correction fitted to three simulated scenarios returns stiffness ratios of 1.55 and 2.53 for reference values of 1.5 and 2.63 in the imposed time-series test. This correction has not been validated across other cell geometries or independent RI phantoms.

\subsection{Precision under realistic camera noise}
\begin{figure}[t]
\centering
\includegraphics[width=\columnwidth]{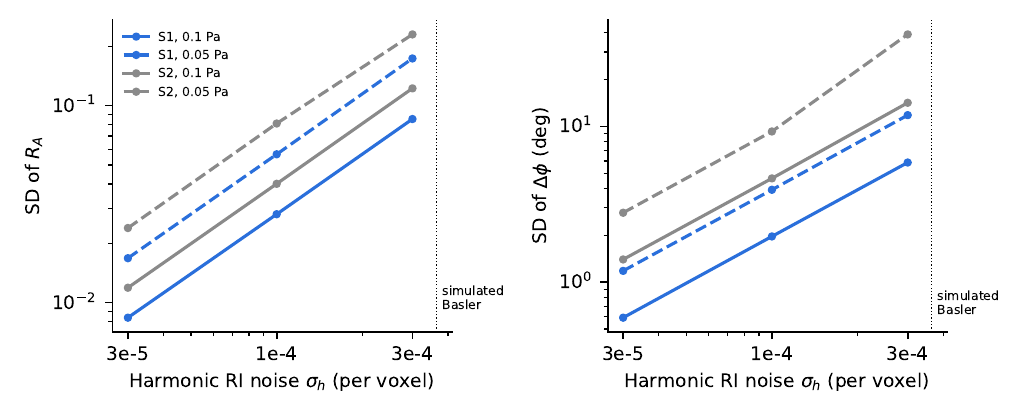}
\caption{Standard deviation of recovered $\RA$ and $\Dphi$ versus first-harmonic RI noise for two stiffness scenarios at 0.1 and 0.05 Pa traction (100 realizations each). Dotted line: single-acquisition noise of the camera-level simulation.}
\label{fig:noise}
\end{figure}

With a first-harmonic RI noise of $\sigma_h=10^{-4}$ per voxel and 0.1 Pa traction, $\RA$ is recovered with a standard deviation of 0.028 to 0.040 and $\Dphi$ with $2.0^\circ$-$4.6^\circ$ over 100 independent noise realizations; S1 and S2 are separated by about 7 standard deviations (Fig.~\ref{fig:noise}). Standard deviations scale with $\sigma_h/\tau$ and decrease with the square root of the number of averaged acquisitions.

The camera-level simulation places one 100-angle, 16-visit acquisition (5.3 s) at $\sigma_h\approx3.6\times10^{-4}$. In this regime $\RA$ is recovered from a single acquisition with an uncertainty of 0.10 to 0.15, which separates the two stiffness scenarios at about $2\sigma$, while averaging several acquisitions brings both observables to the precision required for within-cell tracking (Table~\ref{tab:targets}).

\begin{table}[t]
\caption{Acquisition requirements at 0.1 Pa traction ($\sigma_h=3.6\times10^{-4}$ per acquisition, $3\sigma$ criterion)}
\label{tab:targets}
\centering
\footnotesize
\setlength{\tabcolsep}{3pt}
\begin{tabular}{L{0.44\columnwidth}L{0.46\columnwidth}}
\toprule
Measurement goal & Requirement \\
\midrule
Stiffness contrast between cells ($r_G$ 1.5 vs 5) & 2 to 3 acquisitions ($\approx$10 to 15 s) \\
Within-cell MAR ($\Delta\RA\approx0.22$, $3\sigma$) & 4 to 6 acquisitions per time point, or $\approx$0.18 Pa \\
Phase contrast $\Dphi$ to $\pm2^\circ$ & $\approx$13 acquisitions, or $\approx$0.36 Pa, or $\sigma_h\le10^{-4}$ \\
\bottomrule
\end{tabular}
\end{table}

\subsection{Tracking stiffening and recovery on a single cell}
\begin{figure}[t]
\centering
\includegraphics[width=\columnwidth]{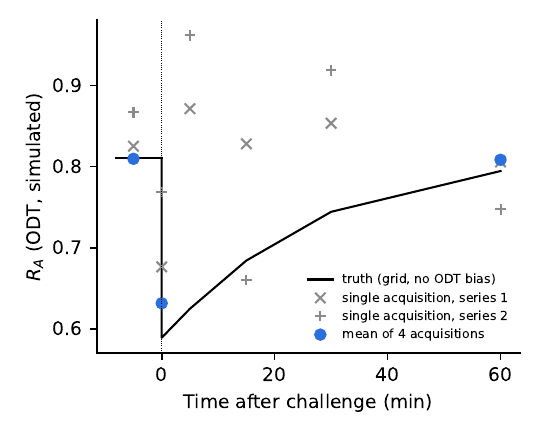}
\caption{Readout of an imposed nuclear stiffening and recovery. Line: reference $\RA$; grey markers: single simulated acquisitions; blue markers: means of four acquisitions.}
\label{fig:recovery}
\end{figure}

To test the adaptation readout, the nuclear stiffness of the reference cell was stepped up by 75\%, a magnitude reported for isolated nuclei after low-intensity vibration \cite{Newberg2020}, and allowed to recover exponentially with a 20 min time constant. The reference $\RA$ falls from 0.81 to 0.59 and returns to 0.79 by 60 min. Averaging four camera-level acquisitions per time point recovers $\RA=0.810$, 0.632, and 0.808 at $-5$, 0, and 60 min, giving $A=0.18$ (reference 0.22), $\rho=0.99$ (0.93), and MAR $=0.18$ (0.21): $\RA$ stays within 7\% of the reference at each time point and MAR within 15\% (Fig.~\ref{fig:recovery}); $A$ is underestimated by 20\% because the ODT bias is larger in the stiffened state. Single acquisitions capture the stiffening step but not the recovery course, confirming that modest averaging is the key enabler for longitudinal measurements.

\section{Experimental Platform: Tomograms and Preliminary Five-Cell Readouts}
\label{sec:exp}
\subsection{Hologram acquisition and reconstruction}
\begin{figure}[t]
\centering
\IfFileExists{figures/E0_raw_hologram.png}{%
  \includegraphics[width=0.9\columnwidth]{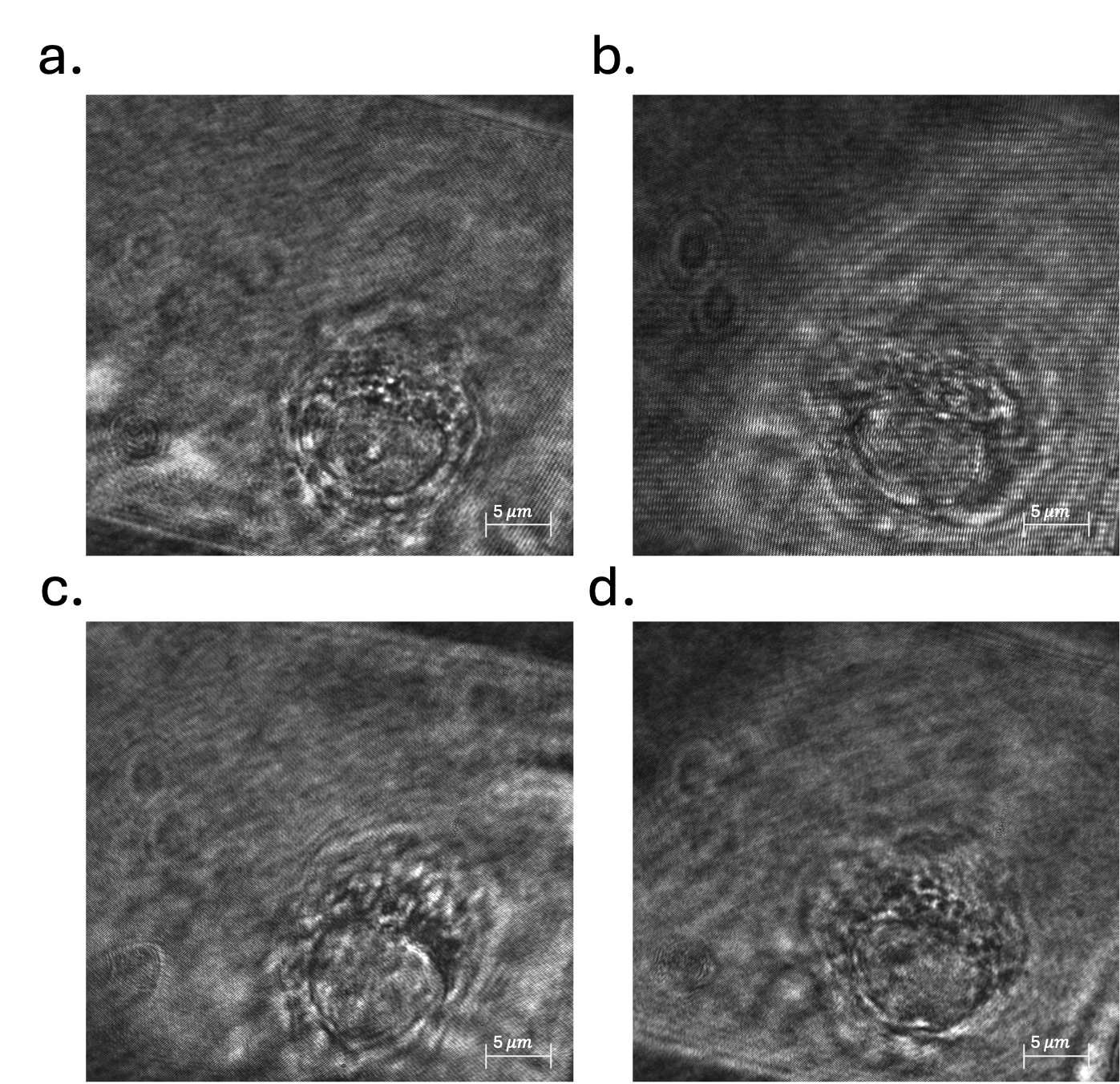}}{%
  \exppanel{2.3cm}{Experimental figure: raw hologram}{Off-axis hologram of an MDA-MB-231 cell.}}
\caption{Off-axis holograms of an MDA-MB-231 cell at four of the 200 illumination angles of one series: (a) 0.38, (b) 0.97, (c) 0.98 and (d) 0.67 of the detection NA, measured from the position of the unscattered component in each hologram spectrum (uncertainty about $\pm0.03$). The fringe pattern and the diffraction pattern of the cell change with the illumination angle.}
\label{fig:exp_a}
\end{figure}

Off-axis holograms are recorded as the DMD varies the illumination angle (Fig.~\ref{fig:exp_a}). The representative static series shown in that figure contains 200 angular illuminations, while the simulated phase-resolved protocol uses 100 angles. Extra angular views increase sampling density but do not, by themselves, extend the accessible numerical aperture. For static reconstruction, the complex field is retrieved by Fourier filtering of the off-axis sideband, and the 3-D RI is reconstructed using the Rytov approximation and Fourier-diffraction (Ewald-sphere) mapping \cite{Wolf1969,Sung2009}.

\subsection{Three-dimensional tomograms of MDA-MB-231 cells}
\begin{figure}[t]
\centering
\IfFileExists{figures/E1_tomogram_cell_A.png}{%
  \includegraphics[width=0.85\columnwidth]{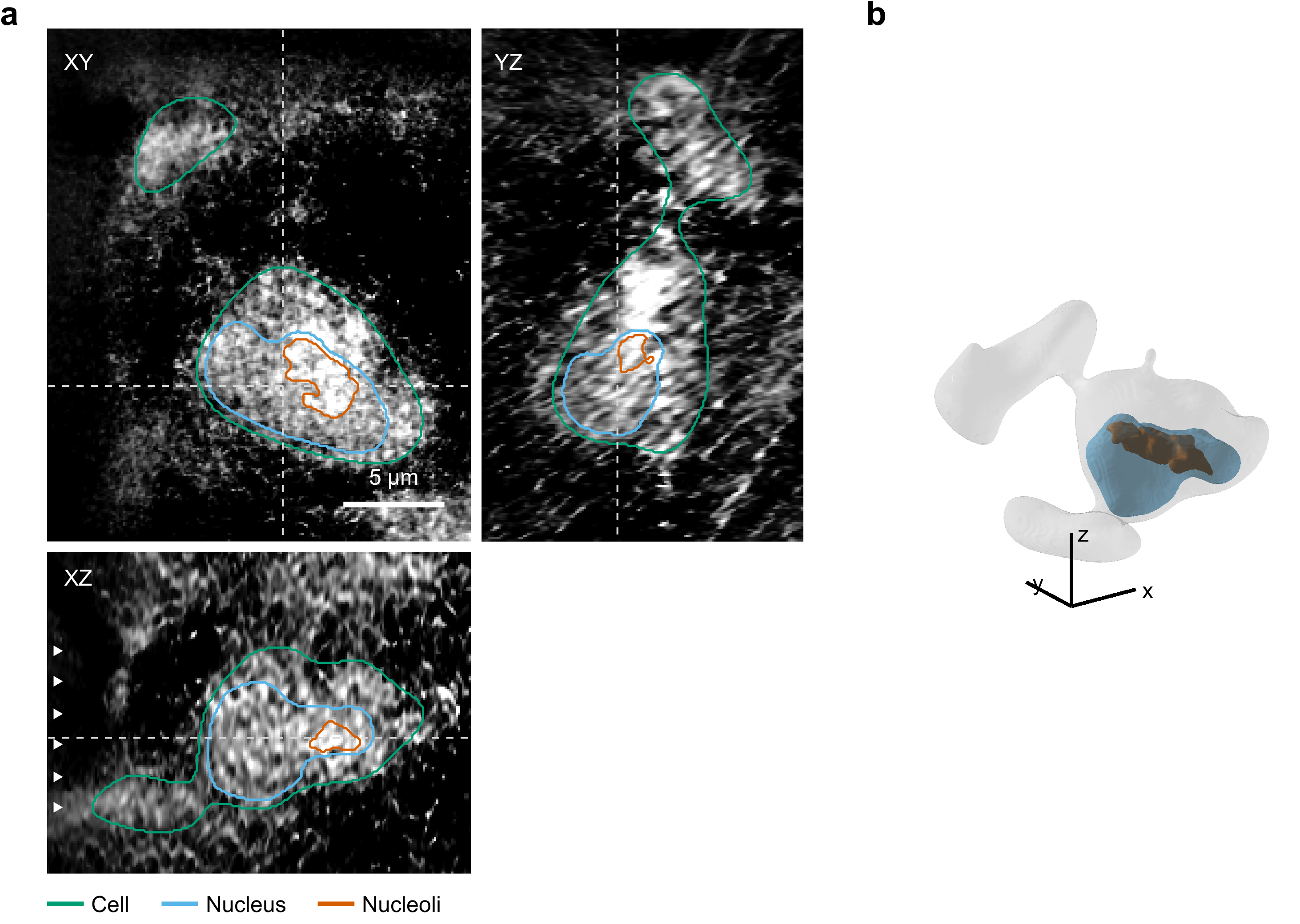}}{%
  \exppanel{2.6cm}{Experimental figure: RI tomogram}{Orthogonal slices with cell, nucleus\\and nucleoli boundaries.}}
\caption{Three-dimensional refractive-index tomogram of an adherent MDA-MB-231 cell: orthogonal XY, YZ and XZ slices with the segmented cell (green), nucleus (blue) and nucleoli (orange) boundaries; scale bar 5 $\mu$m. The assigned nuclear and candidate nucleolar regions are based on RI morphology rather than independent molecular labeling.}
\label{fig:exp_ri}
\end{figure}

Figure~\ref{fig:exp_ri} shows the RI tomogram of an adherent MDA-MB-231 cell. The nucleus and the nucleoli appear as regions of higher RI within the cytoplasm and are segmented in 3-D, which provides the compartments on which $\RA$ and $\Dphi$ are to be evaluated. Without an independent molecular marker, the assignments of nucleolar regions should be considered tentative. Absolute compartment RI values depend on the reconstruction and segmentation settings; they will be reported once the processing pipeline is fixed and validated with reference beads of known RI.

\subsection{Static tomogram differencing}
\begin{figure*}[t]
\centering
\IfFileExists{figures/S1_static_differencing.png}{%
  \includegraphics[width=0.82\textwidth]{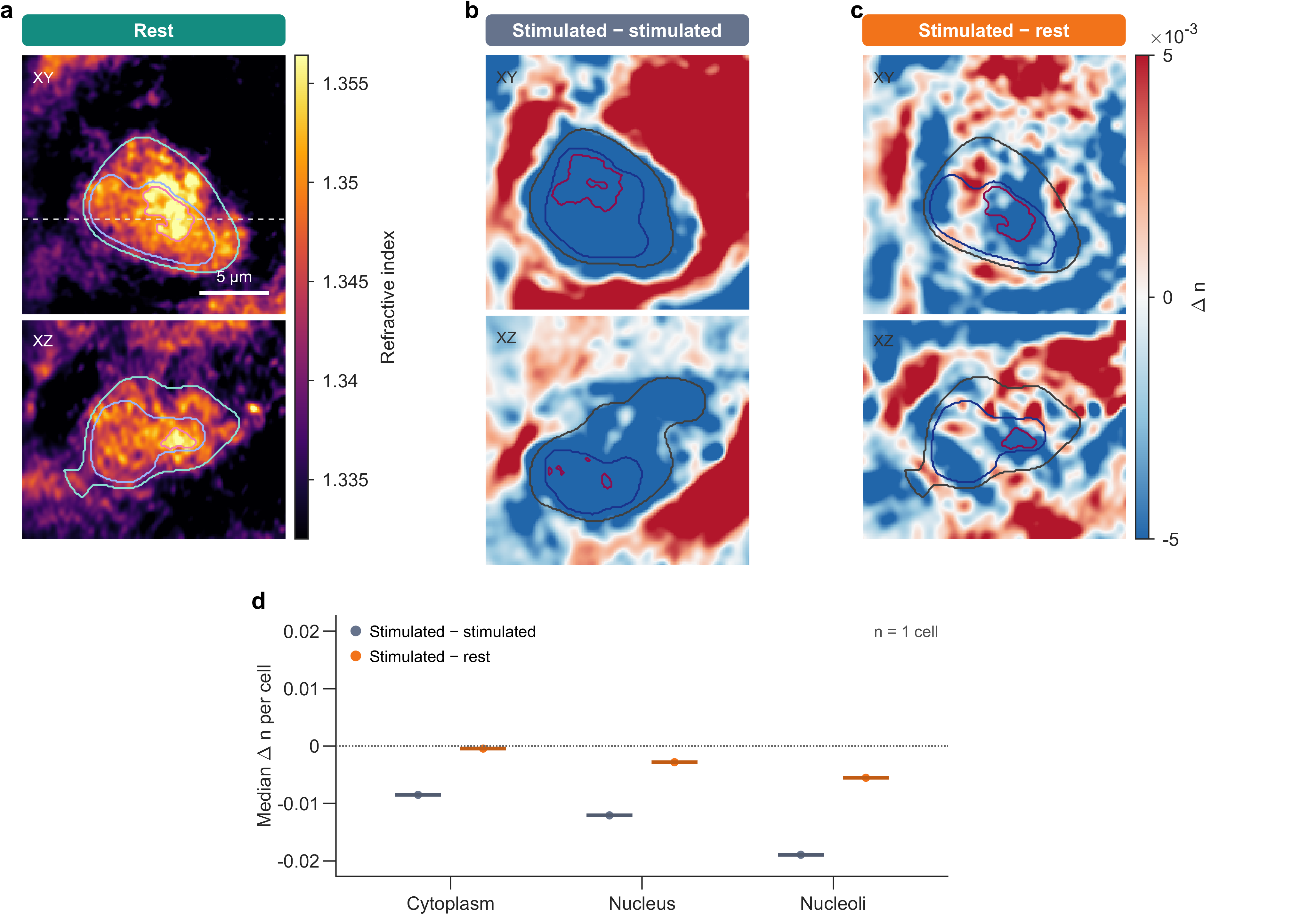}}{%
  \fbox{\parbox[c][3cm][c]{0.8\textwidth}{\centering Figure: static tomogram differencing}}}
\caption{Static tomogram differencing in one MDA-MB-231 cell. (a) Rest tomogram with the cell, nucleus and nucleoli boundaries; dashed line: position of the XZ slice; scale bar 5 $\mu$m. (b) Difference between two acquisitions during stimulation and (c) between stimulation and rest; boundaries of the cell (grey), nucleus (dark blue) and nucleoli (red); the color scale saturates at $\pm5\times10^{-3}$. (d) Median $\Delta n$ per compartment. The same-condition difference exceeds the stimulation-rest difference; one stimulated acquisition was displaced axially by 7.2 $\mu$m. $n=1$ cell.}
\label{fig:static_diff}
\end{figure*}

To test whether acoustic stimulation alters the static RI distribution, separate static tomograms of one MDA-MB-231 cell were compared at rest and during stimulation (Fig.~\ref{fig:static_diff}). The difference between two acquisitions under the same condition ($-0.009$ to $-0.019$ in compartment medians) exceeded the difference between stimulation and rest ($-0.0005$ to $-0.0055$); one of the two stimulated acquisitions was displaced axially by 7.2 $\mu$m relative to the other. Differencing separate static acquisitions is therefore dominated by acquisition-to-acquisition variability such as focus drift and cannot resolve the nanometer-scale response predicted in Section~III. This motivates the phase-locked scheme of Fig.~\ref{fig:workflow}, in which all modulation phases are recorded within a single acquisition at each illumination angle. A stabilized repeat with rest-rest controls on several cells is planned.

\subsection{Preliminary nuclear and cytoplasmic responses in five cells}
Table~\ref{tab:exp5} summarizes the experimental results from five MDA-MB-231 cells. The measured nuclear-to-cytoplasmic amplitude ratios range from 0.50 to 0.73, with a mean of $0.616\pm0.098$ (sample standard deviation). All five values are below one. If the experimental extraction follows the regional shear-response definition in (\ref{eq:metrics}), this indicates a lower estimated nuclear response amplitude than that of the cytoplasm in these cells. The mechanical model also predicts ratios below one for its selected stiffness scenarios, providing a qualitative point of agreement. The measured cells are not matched to specific simulated cases, so no cell-wise simulation error is calculated.

The measured relative phases range from $-2.50^\circ$ to $-1.12^\circ$, with a mean of $-1.91^\circ\pm0.58^\circ$. All five values are negative, as in the simulated scenarios where the nucleus responds later than the cytoplasm. This shared sign is encouraging, but the phase differences are small. Without repeat measurements and an independently measured phase-noise floor, we cannot determine how reliably these lags can be separated from drift or processing artifacts. The observations are therefore presented as preliminary patterns rather than validation of a specific rheological model. To compare experimental and simulated trends quantitatively, both datasets will need a shared independent variable, such as frequency or stiffness contrast.

\begin{table}[t]
\caption{Preliminary nuclear-to-cytoplasmic amplitude ratio $\RA$ and relative phase $\Dphi$ measured in five MDA-MB-231 cells. The final row reports the mean and between-cell sample standard deviation ($n=5$), not repeated-measurement uncertainty.}
\label{tab:exp5}
\centering
\footnotesize
\setlength{\tabcolsep}{10pt}
\begin{tabular}{crr}
\toprule
Cell & $\RA$ & $\Dphi$ ($^\circ$) \\
\midrule
1 & 0.64 & $-1.12$ \\
2 & 0.68 & $-2.31$ \\
3 & 0.73 & $-1.51$ \\
4 & 0.53 & $-2.10$ \\
5 & 0.50 & $-2.50$ \\
\midrule
Mean $\pm$ SD & $0.616\pm0.098$ & $-1.91\pm0.58$ \\
\bottomrule
\end{tabular}
\end{table}

\subsection{Next step: validation and replication}
The next experiments will focus on reproducibility. Angle and phase timing should be recorded alongside ultrasound-off acquisitions and repeat measurements of the same cell. The phase-noise floor is particularly important because the reported nuclear-to-cytoplasmic phase differences are only a few degrees. Near-cell loading will be calibrated independently, for example by phase-locked tracking of PAA beads. A controlled frequency series or an independent measure of stiffness contrast would then allow experimental trends to be compared directly with the model. We will also check heating, acquisition drift and changes in cell morphology, and report all analyzed cells and exclusion criteria. Proposed acceptance criteria include a temperature rise below 0.5 K per acquisition and a verified first-harmonic noise-to-traction ratio below $3\times10^{-3}$ Pa$^{-1}$.

\section{Discussion}
The simulations suggest that phase-locked acoustic ODT could provide information about both nuclear and cytoplasmic mechanics in adherent cells. Under the mechanical conditions tested here, the amplitude ratio $\RA$ changes mainly with the assumed stiffness contrast, while the relative phase $\Dphi$ is sensitive to differences in dissipation. Both are relative measures, which may reduce sensitivity to overall load changes when both regions experience the same forcing. They do not remove the need to characterize how acoustic stress is distributed across a real cell. The simulated recovery example shows how repeated observations could follow an imposed mechanical change, but it is not experimental evidence of recovery in MDA-MB-231 cells. The reported MAR is simply the net recovered $\RA$ difference, so a physiological interpretation requires longitudinal measurements and independent validation.

The modeling points to two practical constraints. First, sufficient acoustic loading must reach the cell without unwanted heating or flow-induced drift. A lateral traction of 0.05 to 0.1 Pa is a useful target within the simulated assumptions, not an experimentally established force in the present chip. Neither the local stress field nor the assumed acoustic energy density has been calibrated in this geometry. Second, the first-harmonic RI response is weak and sensitive to camera noise, missing-angle transfer and reconstruction choices. For perspective, a 5 nm displacement across an illustrative 300 nm RI transition with contrast 0.004 would yield a first-order modulation of about $6.7\times10^{-5}$, below the single-voxel harmonic noise of $3.6\times10^{-4}$ in the camera simulation. Regional fitting may pool information from many voxels, but its gain depends on RI texture and correlated errors; it must not be inferred from voxel count alone.

The static phantom assigns $n_N=1.362$ and $n_C=1.366$, whereas the experimental tomograms show candidate nuclear regions with different local contrasts. This is not necessarily inconsistent, because real RI distributions are heterogeneous, but the inferred motion depends directly on RI gradients. Recovery with other nuclear RI values, heterogeneous boundaries, different segmentation masks and optical missing-cone conditions needs separate testing. We have not yet established the smallest singular values or conditioning of the regional inversion for these cases. Improving phase stability and averaging acquisitions may help, but experimental phase uncertainty and the usable illumination angles must be measured rather than assumed.

Our experimental work provides initial ODT tomograms and five regional measurements. Across these cells, $\RA$ is consistently less than one and $\Dphi$ is negative, following the qualitative direction of selected mechanical simulations. These results are encouraging because they identify response patterns worth testing further. The small cohort and lack of repeated phase measurements prevent us from assigning a quantitative stiffness or damping contrast to the cells. Independent acoustic calibration, validated image segmentation and a controlled series with a shared physical variable are needed before model-predicted trends can be tested directly. A coupled acoustofluidic model of the chip \cite{Bach2018} would also help relate the transducer drive to local mechanical stress. With those steps, the method could be extended toward studies of mechanical adaptation in other cell types.

\section{Conclusions}
We investigated a route from acoustically driven cellular motion to phase-resolved 3-D optical signals and regional nuclear and cytoplasmic responses. The simulated end-to-end chain predicts measurable contrast under specified mechanical loads and camera conditions, and the interleaved timing scheme provides a feasible way to acquire angle- and phase-resolved data. Initial ODT tomograms and five-cell regional measurements motivate further work: the amplitude ratios are below one and the reported nuclear-to-cytoplasmic phase differences are negative, in the directions observed in selected simulated cases. These observations do not yet validate cell-specific moduli or dissipation. The main next steps are a conditioning and recovery analysis of the RI-to-strain inversion, experimental phase-noise controls, repeated measurements and acoustic loading calibration.

\section*{Data and Code Availability}
The data supporting the findings of this study are available from the corresponding author upon reasonable request.

\section*{Acknowledgment}
The schematic illustrations in Figs.~\ref{fig:concept}, \ref{fig:workflow}, \ref{fig:setup} and \ref{fig:odt} were prepared with the assistance of an AI image-generation tool and checked by the authors for accuracy.

\bibliographystyle{IEEEtran}
\bibliography{references}
\end{document}